\documentclass{epl}

\usepackage{graphics}

\title{Characterization of the glass transition in vitreous silica by
temperature scanning small-angle X-ray scattering}

\author{R. Br\"uning\inst{1}\thanks{E-mail: \email{rbruening@mta.ca}} \and C. Levelut\inst{2} \and A. Faivre\inst{2} \and R. LeParc\inst{2}
\and J.-P. Simon\inst{3} \and F. Bley\inst{3} \and J.-L. Hazemann\inst{4}}

\institute{
  \inst{1}Physics Department, Mount Allison University, 67 York Street,
              Sackville (NB), Canada E4L 1E6\\
  \inst{2}Laboratoire des Verres, UMR 5587 CNRS-Universit\'e Montpellier II,
              cc 69, 34095 Montpellier Cedex, France\\
   \inst{3}LTPCM-ENSEEG-INPG, UMR-CNRS 5614, B.P. 75,
              38042 St.\ Martin d'H\`eres Cedex, France\\
  \inst{4}Laboratoire de Cristallographie, UPR 5031, CNRS, B.P. 166, 38042 Grenoble, 
               France\\
}

\shortauthor{R. Br\"uning \etal}
\shorttitle{Characterization of the glass transiton etc.}

 {Glass transition}

\begin{document}
\maketitle

\begin{abstract}
The temperature dependence of the x-ray scattering in the region below
the first sharp diffraction peak was measured for silica glasses with low and
high OH content (GE-124 and Corning 7980).
Data were obtained upon scanning the temperature
at 10, 40 and 80 K/min between 400 K and
1820 K.  The measurements resolve, for the first time,
the hysteresis between heating and cooling through
the glass transition for silica glass, and the data
have a better signal to noise ratio than previous light scattering and
differential thermal analysis data.  For the glass with the
higher hydroxyl concentration the glass transition is broader and at a lower
temperature. 
Fits of the data
to the Adam-Gibbs-Fulcher equation provide updated kinetic parameters
for this very strong glass.
The temperature derivative of the observed X-ray scattering matches
that of light scattering to within 14\%.
\end{abstract}

\section{Introduction}
Although vitreous silica is of great importance for fiber optics communication, technology
and science,
there has been a relative paucity of data regarding its glass transition.  In overviews of
the properties of glasses, pure vitreous SiO$_2$ had to remain a missing data point \cite{boh}. 
Scanning calorimetry is perhaps the most commonly used
tool to determine the glass transition temperature, $T_g$, and the relaxation
state of glasses.  For vitreous silica, due to the high $T_g$, this method cannot be used
without some difficulty, and the first in situ temperature scanning experiment for this material
monitored the light scattering \cite{sai1}.   More
recently, differential thermal analysis (DTA) measurements have determined the step
in the specific heat at the glass transition.  An unusual endothermic process takes place
during annealing \cite{bru4}, which can be attributed to the diffusion of water/hydroxyl
groups \cite{tom}.  The response of vitreous silica to quenching at different rates has
been investigated by a molecular dynamics simulation \cite{vol}.
High temperature small-angle X-ray scattering (SAXS) experiments have been
proven as a sensitive method to detect the glass transition, and changes of the state of the
glass \cite{lev,lep}.  The relation between thermal analysis methods and
temperature scanning SAXS has been established \cite{fai}, and
here we use this method to obtain reference
data of high quality for the glass transition in vitreous silica.

\section{Experimental Method}
SAXS measurements were carried
with the D2AM experimental set-up at the European Synchrotron Radiation Facility (ESRF)
at a wavelength of $\lambda = 0.677 \, \rm \AA$.
Polished plates of GE-124 (six samples) and Corning 7980 (three samples) were measured.
The hydroxyl content of the samples was determined with infrared spectroscopy.
For the GE-124 samples, $\rm [OH] = (2 \pm 1)$ wt.\  ppm, did not change
during the experiment. For Corning 7980, the samples in the as-received
state  contained $(900 \pm 6)$ wt.\  ppm, and the average [OH] decreased to $(832 \pm 6)$  wt.\  ppm in the course of the  measurements.
Samples, with dimensions of $1 \times 4 \times 12$ mm$^3$,
were placed inside a molybdenum furnace
equipped with Kapton windows \cite{sol,lep,lev}.
During the SAXS transmission mode
measurements the temperature was scanned at rates of 10, 40 and 80 K/min,
while helium near atmospheric pressure flowed through the furnace.
The observed melting temperature of a gold foil upon heating at 40 K/min
deviated 2 K from the standard value.  The heating and cooling
curves obtained with the vitreous silica samples are consistent,
so that thermal gradients between samples and the thermocouples are negligible.
The thermal history of the samples was reset by heating to the top of the temperature range, followed
by cooling and heating at the desired rate.  For each sample
the cool-heat cycle was repeated, typically three
times. At the sample surface a small amount of crystallization is visible after the measurements,
but no correlation between the SAXS signal and the progress of surface crystallization was found.
The SAXS was observed with a fiber optic
coupled CCD camera.  The acquisition time was 20 s (10 s for 80 K/min scans).
Distortions and cosmic tracks in the CCD signal were corrected before
integrating it in the azimuthal direction.
The x-ray was monitored before the furnace, $m_0$, and after the furnace, $m_1$,
 by scattering part of the beam
with a Kapton foils onto scintillation counters. Between sample changes the SAXS signal of the empty
furnace, $I^B(q)$, was measured.  Here $q = 4 \pi \lambda^{-1} \sin \theta$ is the scattering vector,
and $\theta$
is half of the scattering angle.  This background originated mainly from the Kapton windows and air scattering.
Based on the measured signal with the sample in place,
$I^{S+B}(q)$, the signal originating from the sample was calculated according to
\begin{equation}
I(q) \propto  (t \ln 1/t)^{-1} [I^{S+B}(q)/m_1^{S+B} - I^{B}(q)/m_1^{B}] m_1^{S+B} / m_0^{S+B},
\end{equation}
where the transmission factor $t = m_1^{S+B} m_0^{B} / (m_0^{S+B}  m_1^{B})$
allows for changes in the thickness and orientation of the sample.
The scattered intensity was converted to absolute values,
in electron units per SiO$_2$ composition unit, by scaling the data to the measured signal obtained with water.
We assumed the value ofÊ 6.37 e.u.\ per
H$_2$O molecule, which is approximately constant
from $q = 0.3$ to $0.7 \, \rm \AA^{-1}$ \cite {am_lev}.
The probable error of the conversion factor is 4\%.

The analysis of the data follows the procedure described in \cite{lev}.
\begin{figure}[tbp]
\includegraphics{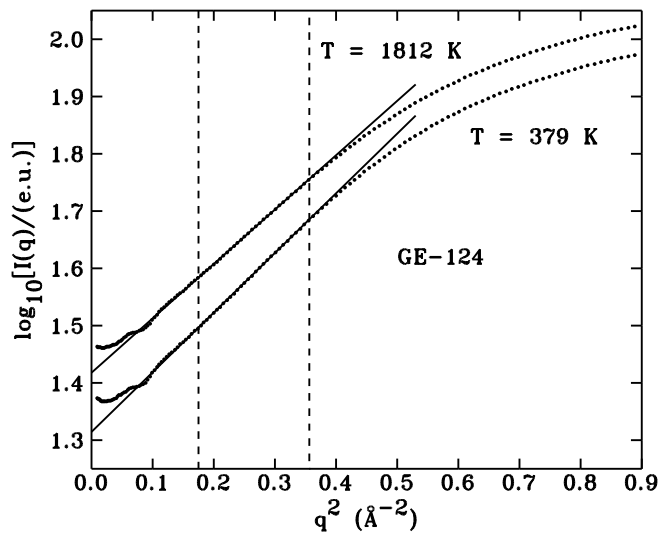}
\caption{\label{fig1}
Logarithm of the scatted intensity of GE-124 plotted versus $q^2$
at two temperatures.  The solid lines are obtained by linear regression
of the data between the dashed vertical lines, and $I(q=0)$ is obtained
as one of the regression parameters.
}
\end{figure}
Fig.\ \ref{fig1} shows $I(q)$, on a logarithmic scale, as a function of $q^2$.
Most of the data range corresponds to the low-$q$ tail of the first sharp
diffraction peak (FSDP) of vitreous silica \cite{bru}.
The small structure in the spectrum around $q^2 = 0.07 \rm \, \AA^{-2}$
is due to residual scattering originating from the Kapton windows.  Precision SAXS 
data show that the SAXS intensity in vitreous silica continues to decrease down
to $q = 0.01 \,{\rm \AA^{-1}}$ \cite{am_lev,ren}, so that the rise observed here
at low angles is not an inherent property of the material.
The linear regime in this plot, extending from
$q^2 \approx 0.04$ to 0.4 $\rm \AA^{-2}$, is well
described by $I(q) = I(0) \exp (bq^2)$. 
For each temperature, the parameters $b$ and $I(0)$ are obtained by
linear regression.
The data range included in the regression calculation is indicated by the
broken vertical lines in fig.\ \ref{fig1}. The
regression parameters depend
only weakly on the range of data.

\begin{figure}[tbp]
\includegraphics{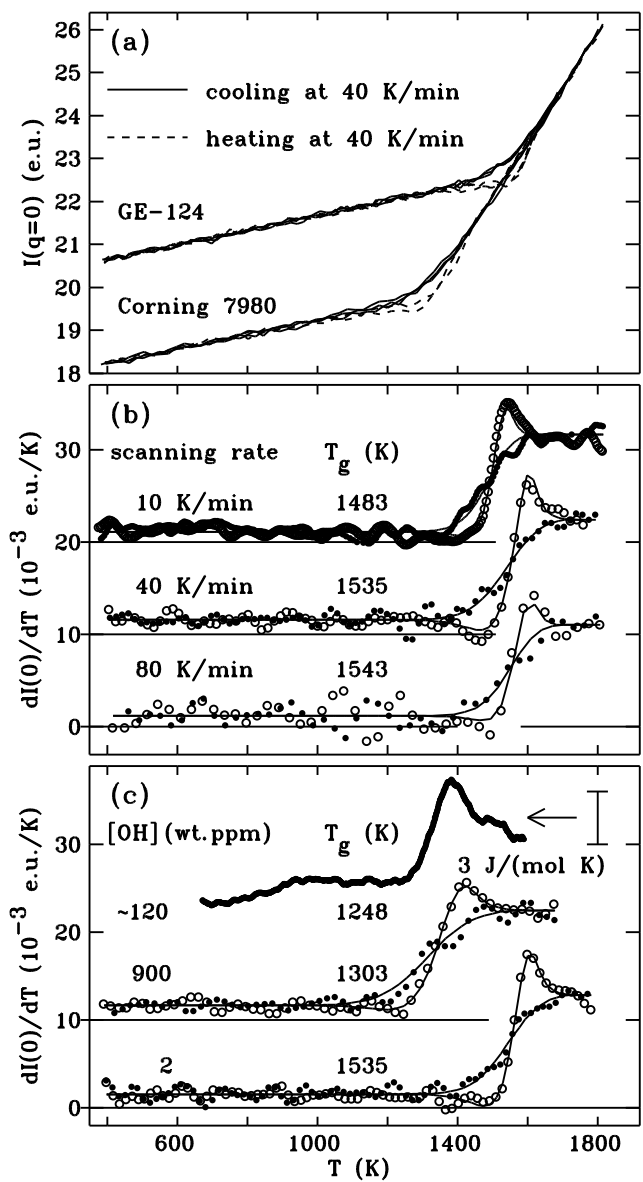}
\caption{\label{fig2}
(a) Scattered intensity, extrapolated to $q=0$, as a function of
temperature.
GE-124 contains $2$ wt.\ ppm H$_2$O, and Corning 7980 contains
$900$ wt.\  ppm H$_2$O.
(b) Temperature derivatives of $I(0)$, obtained for GE-124 upon scanning
the temperature at different rates.
Full and open symbols show cooling and heating data, respectively.
Solid lines are fits to the AGF equation. For clarity, the 40 and 10 K/min
curves were
shifted up by 10 and $20\times 10^{-3}$ e.u./K, respectively.
(c) Temperature derivatives of $I(0)$ data obtained at
40 K/min  for GE-124 and,
shifted up by $10\times 10^{-3}$ e.u./K, Corning 7980.  Top: heat flow upon heating a sample
with $\approx 120 \rm \, wt.\ ppm$ at 40 K/min (ref.\  \cite{bru4}).
}
\end{figure}

The extrapolated intensity, $I(0)$, rises with increasing temperature, while
the slope $b$ decreases (fig.\ \ref{fig1}).
This temperature dependence is
qualitatively similar to results obtained with a fragile glass \cite{fai}.
For brevity we restrict the discussion to $I(0)$,
which has lower experimental uncertainty.
For GE-124, the overall change of $I(0)$
is about 20\% over the whole temperature range of about 1400 K, fig.\ \ref{fig2}(a).
In the successive temperature scans on the same sample,
$I(0)$ changes slowly by about 2\%, and, once upon a refill of the synchrotron,
$I(0)$ increased by
3\%.  Hence we attribute these effects to changes of the incident beam
(e.g. of the energy, the direction or position) which are not fully eliminated
by normalizing the signal to the measured X-ray flux before the sample.
Linear corrections (less than 2.5\%) were applied to the traces shown in fig.\ \ref{fig2}(a)
in order to align the end points of the traces obtained with a given sample .

\section{Results and Discussion}

The temperature dependence of the SAXS in
fig.\ \ref{fig2}(a) shows the standard features of the glass
transition: $I(0)$ changes gradually
with temperature in the glass state, whereas
the supercooled liquid state depends more strongly on
temperature.  In addition to
the change of slope, the glass transition is marked by a hysteresis between
the heating and cooling curves.
The $T_g$ of Corning 7890 vitreous silica, with a higher water content,
is ($232 \pm 14$) K lower than the $T_g$ of GE-124.
Outside the glass transition
regime the slopes of the curves are constant.  The results of multiple
measurements on a given sample, as well as on equivalent samples, agree.
A slight difference can be seen in the heating curves
for the water-rich sample, which may be caused by the
partial loss of water at the sample surface.

The slopes of the straight line segments of the data in fig.\ \ref{fig2}(a) are given
in table \ref{slopes},
as well as the corresponding slopes obtained by light scattering \cite{sai1}.
The slopes are normalized to the intensity at
an arbitrary temperature above $T_g$, where the samples are in metastable
equilibrium (1800 K).  The light scattering based slopes are higher
than the SAXS based results by 11\% above
$T_g$, and 14\% below $T_g$.
Since the data are based on very different
ranges of photon momentum transfer (extrapolation to $q= 0$ from the range
$0.2 \, {\rm \AA^{-1}} < q < 0.6 \,{\rm \AA^{-1}}$ for SAXS,
and $q =1.8 \times 10^{-3} \,{\rm \AA^{-1}}$ for the light scattering experiment), the
approximate agreement of the x-ray and light scattering results is reassuring.  
It shows that the temperature dependence of density
fluctuations is almost wavelength--independent below as well as above $T_g$.
In one-component materials,
the $q=0$ limit is related to the compressibility \cite{gui}.
Low--$q$ scattering in vitreous silica originates predominantly
density fluctuations, since the chemical order of the SiO$_4$ tetrahedra is
nearly perfect \cite{sai1,lev}.  Following the analysis of the light
scattering data, the slope below  $T_g$ is proportional to the adiabatic compressibility of
the glass, and the slope above $T_g$ is proportional to the isothermal compressibility of the
liquid \cite{sch,sai1}.  A review of literature data for
SAXS and light scattering measurements shows that the compressibilities
determined by different methods vary by up to 25\% \cite{wat}.

\begin{table}[t]
\caption{\label{slopes}
Comparison of the temperature sensitivity of SAXS and light scattering (last two columns).
Normalized temperature derivatives,
$[I(1800 {\rm \, K})]^{-1} dI/dT$,
in units of $10^{-6} \, \rm K^{-1}$,
are evaluated below and above the glass transition regime.  The samples are GE-124, Corning 7980,
and, from ref.\ \cite{sai1}, type A and type G vitreous silica.
}
\begin{tabular}{c||c|c|c|c}
[OH]  (wt.\ ppm)& $2$ & 900 & $<1$  & 1200\\
\hline
$T< T_g$  &    $\ 74 \pm 2$ & \  $72 \pm 2$ & \ $85 \pm 2$ & \ $85 \pm 3$ \\
$T> T_g$  & $500 \pm 6$  & $500 \pm 6$ & $561 \pm 4$ & $561 \pm 4$\\
\end{tabular}
\end{table}

As a next step we consider the temperature derivative of $I(0)$,
which provides curves that are comparable to e.g.\ calorimetric measurements.
While linear adjustments were made for the traces in fig.\ \ref{fig2}(a) (as mentioned above),
this was not necessary for fig.\ \ref{fig2}(b) and (c), since differentiation reduces
the effect of the slow drift of $I(0)$.
High frequencies
are enhanced by numeric differentiation.  These were partially suppressed by applying
a Fourier filter to the differentiated data.
The results obtained for GE-124 at three different scanning rates are presented in fig.\ \ref{fig2}(b).
The differentiated curves clearly show the expected hysteresis of the glass state
between heating and cooling.  The value of
$T_g$ considered here is the low temperature limit of the fictive temperature calculated from the
cooling curve of the best fit of the data to the AGF equation (described below).
This value is close to the onset of the glass transition,
but it has the advantage that it is based on the complete cooling and heating trace.
The fitted value of $T_g$ varies by about 4 K between repeated scans for GE-124, and
by 12 K for Corning 7980.  The glass transition shifts to
higher temperature with increasing scanning rate.  A
Kissinger plot  of $\ln (\phi T_g^{-2})$ vs.\ $T_g^{-1}$, where $\phi$ is the temperature
scanning rate,
yields an activation energy of $E_a/k_B = (62 \pm 10) \times 10^3 \rm \, K$ for GE-124, in
agreement with the slope of viscosity data in an Arrhenius plot, $61.9 \times 10^3 \rm \, K$
for a vitreous silica with [OH] $< 10$ wt.\ ppm \cite{urb}. 

Fig.\ \ref{fig2}(c) shows the differentiated data for the GE-124 and Corning 7980 glasses, as well
as previously published DTA data obtained for
annealed Corning 7980 glass with a reduced hydroxyl content \cite{bru4}.
(In order to demonstrate the degree of reproducibility, the 40 K/min
data for Corning 7980
in fig.\ \ref{fig2} (b) and (c) are examples from different scans.)
A comparison of the SAXS-based curves shows a broader
transition for the glass with the higher hydroxyl concentration.
This agrees with the slope of the equilibrium viscosity in an Arrhenius
plot, which decreases by 28\% as the hydroxyl content increases
from 3 to 1200 wt.\ ppm \cite{het}.
The increase of the width of the glass transition regime may be caused by the joint action of
two processes: At high temperatures direct bond-breaking and relaxation of the SiO$_2$
network take place, while below 1100 K the site-to-site diffusion
of hydroxyl groups is the faster process \cite{tom}.  The hydroxyl diffusion process, emergant
at low temperatures, can therefore broaden the glass transition curve.
The DTA and SAXS based data in fig.\ \ref{fig2}(c) basically agree, which
confirms the results of the prior series of calorimetric measurements \cite{bru}.
However, the SAXS
data are clearly superior both in terms of the stability and linearity
of the background as well as in
providing data upon heating as well as cooling.  One would expect the glass
transition for the 120 wt.\ ppm sample at a temperature 50 K higher than that of the 900 wt.\ ppm
sample \cite{sai1}.

Next the Adam-Gibbs-Fulcher (AGF) equation is used to extract standard kinetic parameters
from the SAXS data.  Details beyond the present summary of this approach can
be found in \cite{bru4,hod}.  The state of the glass can be
approximated by a fictive temperature, $T_f$.
Above the glass transition, $T_f$ is equal to the temperature, while
upon cooling sufficiently far below the glass transition $T_f$ becomes frozen with $T_f = T_g$.
The AGF approximates the relaxation time of the glass and supercooled liquid states as
\begin{equation}
\tau(T,T_f) = A \exp {Q /  { \left[ T \left( 1-T_0/T_f \right) \right] } },
\end{equation}
where the temperatures $Q$ and $T_0$ are parameters,
and $A$ is the high temperature limit of the relaxation time. 
This equation extends the Vogel--Fulcher--Tammann equation, to which
it reduces in the supercooled liquid regime where $T_f = T$ \cite{hod}.
In this case the relaxation time 
diverges when $T$ reaches $T_0$.
The relaxation time is used with the
Narayanaswamy-Moynihan equation, which desribes how $T_f$
evolves as
\begin{equation}
\label{model}
T_f(T) = T^* + \int_{T^*}^T { \left\{ {1 - \exp{\left[ - \left( \int_{T'}^T
{{\rm d}T'' \over {\phi \tau(T'')}} \right)^\beta \right]} } \right\}
{\rm d}T' } .
\end{equation}
This integral involves
a Kohlrausch exponent, $\beta$, as a further parameter.  Finally,
the SAXS data are fitted by treating the temperature derivative
of $I(0)$ far below and far above $T_g$ as further adjustable parameters.
Then we obtain
\begin{equation}
{{dI(0)}\over{dT}} = {\left.{{dI(0)}\over{dT}}\right|_{T \ll T_g}}
+ \left[ \left( {\left.{{dI(0)}\over{dT}}\right|_{T \gg T_g}} -
{\left.{{dI(0)}\over{dT}}\right|_{T \ll T_g}}\right) {dT_f\over{dT}} \right].
\end{equation}
The solid lines fig.\ \ref{fig2}(b),(c) are least-squares fits of the data to the AGF equation.
The kinetic parameters of the fits
are given in table \ref{fitparams}.
\begin{table}[ht]
\caption{\label{fitparams}
Parameters of least squares fits to the AGF equation of the
SAXS data obtained at 40 K/min for GE-124 and Corning 7980.
Experimental errors are estimated from the fit parameter variations
for multiple scans obtained with the same sample.}
\begin{tabular}{c||c|c}
[OH] (wt.\ ppm)                          & $2$            & 900       \\
\hline
\hline
$Q$ ($10^3$ K)                                 & $49 \pm 5$        & $39\pm 9$       \\
$-\lg (A/{\rm s})$                       & $15 \pm 1$        & $11\pm 1$     \\
$T_0$ ($10^3$ K)                               & $0.30 \pm 0.16$ & $0.05\pm 0.11 $  \\
$\beta$                                  & $0.87 \pm 0.06 $  & $0.76\pm 0.05 $      \\
\end{tabular}
\end{table}

For GE-124 ([OH] $= 2$ wt.\  ppm), the AGF fit parameters clearly indicate the properties
of a very strong glass: $T_0$ is only 20\% of $T_g$, $\beta$ is close to one
as expected \cite{boh}, and the time prefactor, $A \sim 10^{-15} \rm \, s$,
is comparable to the inverse of the
Debye frequency. As a comparison, fits to viscosity data
give $Q = 33.53 \times 10^3$ K and $T_0 = 0.54 \times 10^3$ K \cite{sip}.
The fit parameters are sensitive
to the width and the details of the shape of the glass transition curves.  The broader SAXS curve
for the high [OH] sample is reflected in the lower $Q$ and a longer time $A$.
The change of $Q$ matches the above-mentioned 28\% decrease of the activation energy for the
equilibrium viscosities upon increasing [OH] \cite{het}.
The new SAXS data provide tighter constraints on the fit as well as
a superior signal to noise ratio compared to the DTA
data in fig.\ \ref{fig2} (c), so that the present fit parameters are more reliable
than the values given previously  \cite{bru4}.

\section{Conclusion}
The present SAXS data provide, for the first time, a clear picture of the glass transition
in vitreous silica, including the hysteresis between cooling and heating.
The signal to noise ratio is improved over
previous point-by-point SAXS, temperature scanning
light scattering and DTA measurements.
Scanning the temperature at a rate of 40 K/min,
has provided the best data.   The shift of $T_g$ with heating rate matches
the temperature dependence of the viscosity.  Increasing the hydroxyl concentration
from $2$ to $900 \rm \, wt.\ ppm$ lowers $T_g$ by $(232 \pm 14)$ K. The glass with the
higher hydroxyl concentration has a broader glass transition, in agreement with the
results of non-equilibrium viscosity measurements.  The SAXS of vitreous silica
is slightly less temperature
dependent than the light scattering (11\% above and 14\%
below the glass transition).  Based on the present
results, the AGF parameters for vitreous silica have been updated.

\acknowledgements
We wish to thank O. Geyamond, S. Arnaud, and B. Caillot for technical assistance,
as well as J.-F. B\'erar for assistance in using beamline BM02.
ESRF provided the synchrotron radiation facilities, and
the Natural Science and Engineering Research Council
of Canada has supported this work through a Discovery Grant.

\end{document}